\documentclass[letterpaper, 10 pt, conference]{ieeeconf}  
\usepackage[T1]{fontenc}
\usepackage[utf8]{inputenc}
\usepackage{textgreek}

\IEEEoverridecommandlockouts                              

\usepackage{graphicx}
\usepackage{float}
\usepackage{placeins}

\title{\LARGE \bf
Calming Robot Pitches? Exploring the Influence of Robot Voice Pitch on Children’s Stress Levels
}

\author{Nina G. M. van Roij$^{1,2,*}$, Emilia I. Barakova$^{2}$, Briana Isailă$^{3}$, and Aoju Chen $^{1}$
\thanks{$^{1}$Department of Languages, Literature and Communication, Utrecht University, 3512 JK Utrecht, The Netherlands
}%
\thanks{$^{2}$Department of Industrial Design, Eindhoven University of Technology, 5612 AZ Eindhoven, The Netherlands
}%
\thanks{$^{3}$Department of Mathematics and Computer Science, Eindhoven University of Technology, 5612 AZ Eindhoven, The Netherlands
}%
\thanks{$^{*}$Corresponding Author: n.g.m.v.roij@tue.nl
}%
}

\begin{document}
\section*{IEEE Copyright Notice}

This paper has been accepted for publication in the Proceedings of 35th IEEE International Conference on Robot and Human Interactive Communication (RO-MAN 2026). © 2026 IEEE. Personal use of this material is permitted. Permission from IEEE must be obtained for all other uses, in any current or future media, including reprinting/republishing this material for advertising or promotional purposes, creating new collective works, for resale or redistribution to servers or lists, or reuse of any copyrighted component of this work in other works.

\maketitle
\thispagestyle{empty}
\pagestyle{empty}


\begin{abstract}
This study examined whether variations in robot speech pitch influence children’s stress levels during a robot-guided game. Although lower-pitched voices have been shown to facilitate stress regulation in human communication, it remains unclear whether this effect generalizes to synthetic voices in child-robot interactions. Twenty-seven Dutch children aged 8-12 years were randomly assigned to interact with a Zenbo Junior II robot using either a lower-pitched or a higher-pitched voice. The interaction consisted of an introduction followed by a timed LEGO-building game. Stress levels, measured with an adapted version of CAM-S \cite{ekmh13}, increased during the game, confirming the stress-inducing nature of the task. No differences emerged between pitch conditions. These findings suggest that the benefits of lower pitch in reducing stress may not directly translate to child-robot interactions. Possible explanations include children’s developing sensitivity to emotional tone, mismatches between the robot’s voice and appearance, or the use of fixed pitch changes that sound unnatural, since real speech varies dynamically across multiple dimensions. Future research examining combinations of prosodic cues (beyond pitch alone) could provide further insights and help inform robot voice design for effective stress regulation support for children.

\end{abstract}

\section{Introduction}
\label{sec:intro}
Stress is a common experience across developmental stages and life contexts. While moderate levels of stress can facilitate learning and adaptive functioning, chronic or excessive stress negatively affects children's cognitive and emotional development \cite{whiting2021stress}. Children in particular face unique challenges in stress regulation as their prefrontal cortex is still developing and they have had fewer opportunities to develop coping strategies due to their limited life experience\cite{cjdwbgw13}. Early-life stress can therefore have long-term consequences, including cognitive impairments, altered emotion recognition, and difficulties in emotion regulation \cite{pp10}. Early interventions are therefore crucial in mitigating the effects (e.g., \cite{rths19}). 

As technology becomes increasingly integrated into therapeutic and educational environments, social robots have emerged as promising tools for supporting children's stress regulation \cite{lslyhb25}. Their predictable and controllable nature can foster a sense of safety and emotional stability in unfamiliar and anxiety inducing environments \cite{cjaa13}, \cite{n25}. Although their effectiveness has been demonstrated across contexts such as healthcare and education (e.g., \cite{vlfsb24}, \cite{mmkd24}), the mechanisms underlying these effects remain understudied. Robot voice design, with careful attention to prosody could prove to be impactful.

In human communication, prosody plays an crucial role in regulating stress and emotion. For instance, during moments of high tension, therapists often employ a lower and more stable pitch to convey empathy, calmness, and emotional stability \cite{kl06}, \cite{xbviagn14}, \cite{btmrspf10}. Grounding robot voice design in well-established human communication principles may enhance the effectiveness of stress regulation support. For example, applying pitch synchrony in robot interactions \cite{hk18} has been shown to yield similar outcomes to entrainment in human interactions, fostering emotional connection and relaxation \cite{gblh14}.

The present study aims to contribute to the design of more effective and emotionally intelligent social robots, tailored to the needs of children in stress-inducing settings, by examining how robot voice pitch influences stress regulation in children. The following research question was formulated:
\emph{To what extent does manipulating the pitch of a robot’s voice influence children’s stress levels in child-robot interactions?}

Based on empirical findings from human communication (e.g., \cite{btmrspf10}, \cite{kl06}, \cite{tlcl22}), we hypothesize that during a stressful interaction, children interacting with a robot employing a lower-pitched voice will experience lower stress levels than those interaction with a higher-pitched robot voice. 

\section{Related work}
\subsection{ Robot Stress Relief Interventions}
Social robots have demonstrated considerable potential as tools for stress regulation. In acute stressful situations, these systems can provide reassurance, empathetic responses, and calming interactions, thereby supporting users’ emotional well-being \cite{jaapbp23}. Furthermore, social robots have been shown to facilitate the development of sustainable coping strategies by teaching evidence-based stress management techniques \cite{rka23}. For instance, a home companion robot that monitors user distress and delivers personalized interventions has been shown to improve daily functioning \cite{mmkd24}.

Evidence suggests that these benefits extend to children. For instance, in pediatric healthcare, social robots help mitigate stress related to medical procedures and unfamiliar environments. They can provide socio-emotional scaffolding and engaging distractions \cite{jblw18}, \cite{vsbgbc17}, while their predictable interaction style can enhance children's sense of control \cite{cjaa13}. Robot play interactions in emergency departments have been shown to reduce children's stress more effectively than interacting with a nurse or waiting with parents \cite{rsdmvfratb21}. Similar advantages have been reported in mental healthcare settings \cite{n25}. Beyond healthcare, robots have been used to reduce stress during socially demanding tasks by functioning as a communication buffer; for example, in education, teleoperated robot interactions have been shown to lower stress compared to face-to-face instruction \cite{nwyoi15}.

Despite these promising findings, several challenges remain. Parents have reported increased social anxiety in their children during robot interactions \cite{kcmbk21}. Although such observations may partly reflect parental biases, they nevertheless indicate that robot-mediated interactions can introduce unfamiliar social dynamics. Additionally, robot assistance during tasks may reduce children's sense of achievement and perceived self-competence \cite{kcmbk21}. Collectively, these findings suggest that, despite the promise of robot-based stress support, a deeper understanding of the mechanisms underlying child-robot interactions is needed to optimize effectiveness.

\subsection{ The Effect of the Design of a Social Robot on Stress}
A robot's appearance plays an important role in shaping user perceptions and interaction outcomes. For example, more anthropomorphic robots tend to be viewed as more competent and comforting \cite{bcsf21}, and physically embodied robots are more effective in reducing emotional arousal than virtual agents \cite{kra24}. These findings suggest that robot appearance should be a design priority for developing robots to supporting children's stress regulation. 

While visual and embodied aspects of robots design have been extensively studied, voice characteristics have received less attention \cite{htk26}. Yet, voice design strongly influences the perception of personality, competence, and emotional state. For instance, lower-pitched robot voices have been associated with higher perceived interaction quality, whereas higher-pitched voices are often judged as more appealing and attractive \cite{ndnls13}. In child-robot interactions, using child-like voices can increase learning motivation \cite{ws22}, and a robot's voice can shape emotional experiences during storytelling \cite{cder21}. However, findings are mixed. Some studies report no effects of voice manipulation on learning outcomes \cite{mfpbc21}, suggesting factors such as task type, interaction goal, user experience, and voice manipulation type moderate these effects.

Prior research on stress-related applications has explored the influence of voice naturalness and affective expressiveness. Natural female voices have been shown to enhance relaxation in mindfulness contexts compared to synthetic or male voices \cite{mc22}. Similarly, among children with autism spectrum disorder, the use of natural instead of monotonic intonation has been shown to increase positive affect and engagement \cite{vsbgbc17}. Beyond these static voice characteristics, research highlights the importance of dynamic prosodic adaptations. As outlined in Section~\ref{sec:intro}, implementing entrainment, alignment of speech features between interlocutors in human communication \cite{gblh14}, in robot interactions fosters connection and reduces emotional arousal \cite{hk18}. This could suggest that human communication principles may generalize to robot-mediated interactions and may be leveraged to enhance stress support. 

\subsection{The Role of Pitch on stress in human communication}
\label{sec:pitchstressac}
 In human communication, pitch influences the regulation of emotional arousal. Clinical studies show that greater pitch variability in therapists can increase arousal and avoidance behaviors in patients \cite{ww20}, \cite{wfebhww20}, while more stable pitches are associated with increased perceived empathy and competence \cite{tlcl22}, \cite{btmrspf10}. Similarly, lower voice pitch has also been linked to reduced physiological stress \cite{kl06} and greater perceived empathy \cite{xbviagn14}. In addition, patients describe ideal therapists' voices as deeper and lower-pitched \cite{btmrspf10}. These findings may reflect that lower pitch signals emotional stability, as pitch naturally increases during moments of tension \cite{dhod19}. However, pitch manipulations may be used dynamically and interpretation of pitch cues can be context dependent, as research shows that experienced therapists strategically increase pitch during high-stress moments to convey engagement \cite{galhjsh22}. Other prosodic features, such as speaking-rate and intensity, also influence stress responses (e.g., \cite{sjkkmd23}, \cite{kl06}, \cite{btmrspf10}, \cite{tlcl22}), although findings of these cues are arguably less consistent. Overall, a lower, stable pitch emerges as a comparatively robust cue for stress regulation.

Children, like adults, are sensitive to prosodic features from an early age \cite{grossmann2010}, \cite{flglcgg22}. Preschoolers can reliably infer emotional states from pitch contours and adjust their behavior accordingly \cite{quam2011}. Nevertheless, the ability to interpret complex emotions from prosody continues to develop in adolescence \cite{flglcgg22}. Studies on pitch and stress regulation in children suggest similar patterns to adults, with lower pitch facilitating stress reduction. Mothers using more restricted prosody have been shown to reduce infants' physiological stress more effectively \cite{kolacz2021}, and caregiver-produced 'shushing' (a low frequency vocalization) decreases infant arousal \cite{moller2019}. 

Taken together, these findings indicate that children not only perceive emotional information from pitch but also have their affective states shaped by it, much like adults. This underscores the potential of using pitch to regulate children’s stress. Designing robot voices based on these principles may enhance the effectiveness of child-robot stress support interventions. However, much remains unknown about whether the stress reducing effects of lower pitch generalize from human to synthetic voices. The present study addresses this gap.

\section{Methods}
\subsection{Participants}
An a priori power analysis using G*Power 3.1.9.7 \cite{felb07} indicated that a minimum of 25 participants was required to detect a medium-sized effect ($\alpha$ = .05, power = 0.80). Accordingly, 28 monolingual Dutch-speaking children aged 8-12 were recruited through a local after-school care center (N = 12) and convenience sampling (N = 16). One participated in the pilot and was excluded from the analysis, resulting in a final sample of 27 children (M = 10.2 years, SD = 1.5). Care center participants were tested onsite, whereas convenience-sampled participants were tested at home.

The participants were randomly assigned to two experimental groups: Group 1 (N = 14; M = 10.1, SD = 1.5) and Group 2 (N = 13; M = 10.2, SD = 1.7). The groups were comparable in age, with minor differences in gender distribution and testing location.

All children received compensation equivalent to €5. The care center participants received a group gift card, promoting inclusiveness (i.e., not disadvantaging the children that did not participate), and home-tested participants received a treat. Ethical approval was obtained from the ethical review board of Eindhoven University of Technology (ERB2025ID174). Written informed consent from guardians and verbal consent of the children were obtained prior to participation.

\subsection{Experimental Design}
\label{sec:expdes}
A child-robot interaction experiment was conducted in a single twenty-minute session with two components: an introduction and a robot-guided game. In the introduction, participants were familiarized with the robot. This was followed by the game phase consisting of a robot-guided timed LEGO building task, inspired by \cite{lslyhb25}. The study employed a between-groups design. Participants were randomly assigned to one of two experimental conditions applied during the game phase: (1) a higher-pitched robot voice, or (2) a lower-pitched robot voice.

\begin{figure}
    \centering
        \includegraphics[width=0.8\linewidth,
            trim=160 60 160 10, clip]{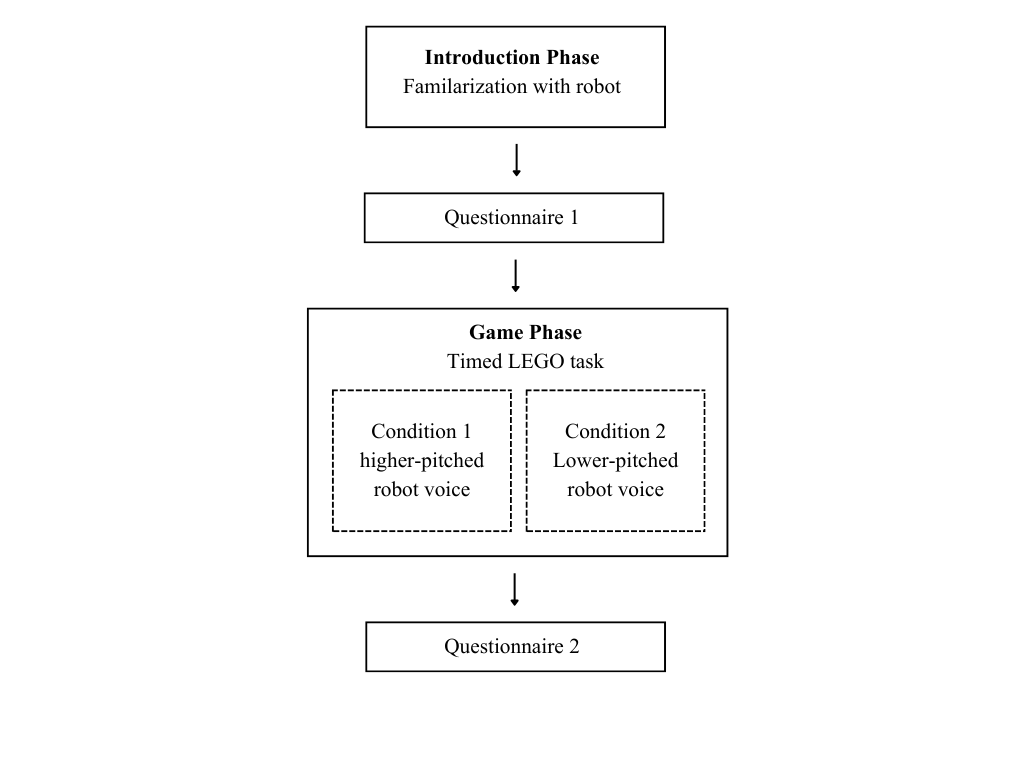}
    \caption{A schematic overview of the experimental design}
    \label{fig:expoverview}
\end{figure}

Voice manipulation served as the predictor variable and children's stress levels as the main outcome variable. Stress was measured using an adapted version of the Children's Anxiety Meter-State (CAM-S) \cite{ekmh13}, translated into Dutch by a trained linguist. With this measure consisting of a 1-10 thermometer-style scale, children indicate their emotional state by coloring the thermometer to reflect how nervous or worried they feel at the moment (Item 1). A second task-focused item was added using the same question format (Item 2) to directly assess the stress associated with the activity and to reduce social desirability biases by increasing psychological distance \cite{se59}, \cite{n07}. Both items were administered twice: once after the introduction (Questionnaire 1) and once after the game (Questionnaire 2). 

\subsection{ Materials}
The materials used in this study were grouped into 3 categories: (1) the robot, (2) the robot voice implementation, (3) the timed LEGO building task. All components are described below.

\subsubsection{The Robot}
\begin{figure}
    \centering
    \includegraphics[width=0.4\linewidth]{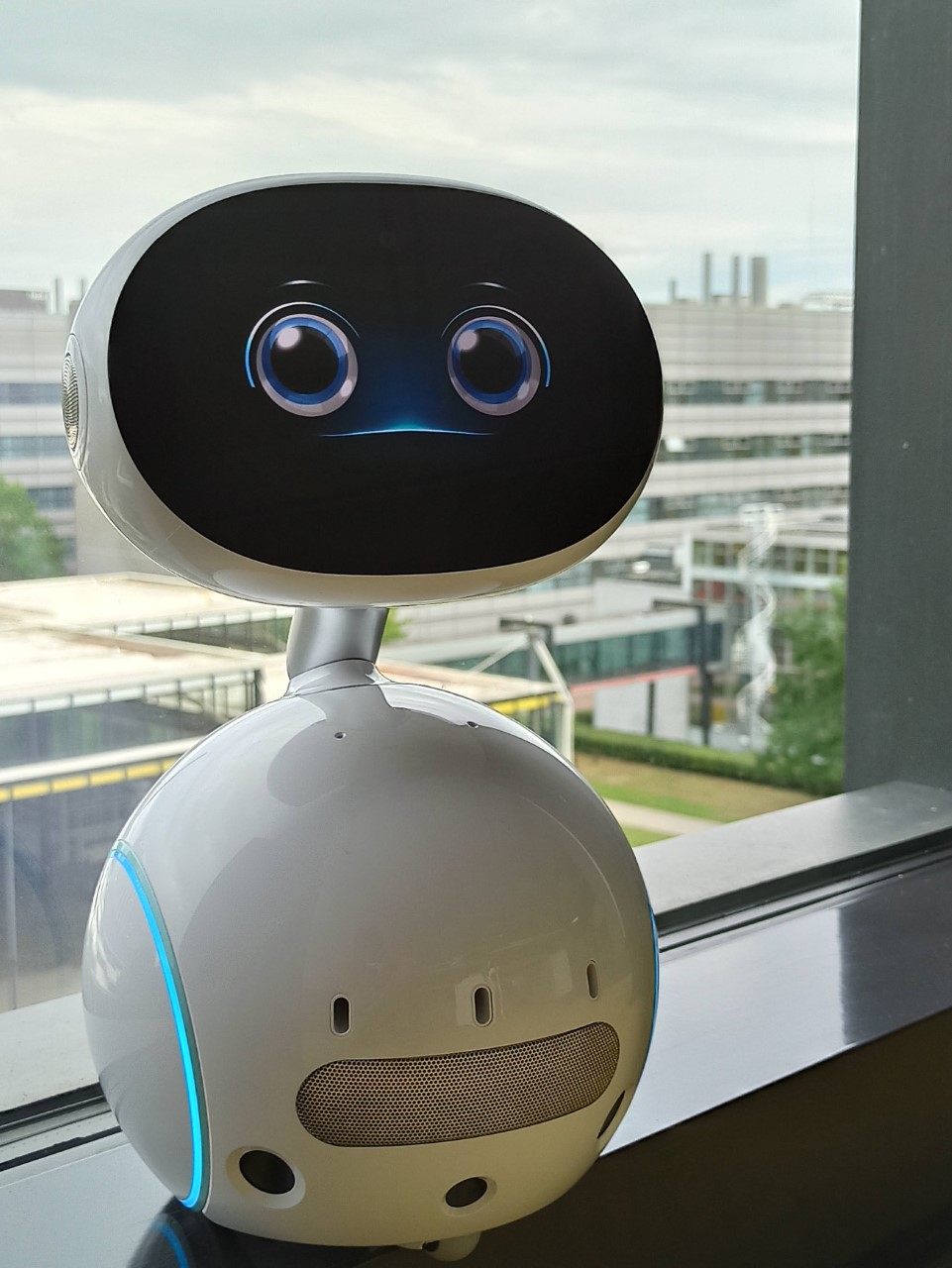}
    \caption{Zenbo Junior II}
    \label{fig:zenbo}
\end{figure}

The experiment used a Zenbo Junior II robot (see Fig.~\ref{fig:zenbo}), equipped with sensors for touch, sound, and vision. The robot's face features a screen-based interface that supports preset and custom facial expressions. In this study, facial tracking enabled the robot to maintain eye contact with participants, while facial presets allowed it to mimic human-like behaviors such as blinking. These features both supported the robot's perception as a socially competent agent \cite{ghbm18, mbdmr18}.

All behavior was fully scripted to ensure consistency between participants. A custom dialogue script of 27 robot speaking turns was developed. This consisted of five turns for the introduction interaction, twenty for the game phase, and two for the closing. To introduce natural linguistic variety and avoid monotony, the game-phase dialogue mixed declarative and imperative sentences. Sentences were intentionally kept short (M = 5.9 words, SD = 1.5) to ensure clarity and comprehensibility, appropriate to the receptive language abilities of the targeted age group.

Because the robot did not come equipped with a preset Dutch voice, audio samples of the dialogue script were generated. Speech synthesis was conducted using Google Cloud Text-to-Speech Engine and a custom python script. A gender-neutral base voice was selected to minimize gender-related biases.

\subsubsection{The Robot Voice Implementation}
\label{sec:robvoiceman}
To create the experimental conditions of the game phase, the fundamental frequency of the synthesized voice was bi-directionally manipulated in PRAAT version 6.4.35 \cite{bw09}. This resulted in three versions: a neutral, unadjusted, voice for the introduction, a higher-pitched voice for Condition 1, and a lower-pitched voice for Condition 2. 

Because this research translates findings from human communication to HRI, a preliminary validation was conducted to confirm that lower-pitched synthetic speech would indeed be perceived as more calming. Supported by previous work showing similar interpretations of emotional prosody in adults and children (see Section~\ref{sec:pitchstressac}), this exploration used adult participants recruited to facilitate online data collection and to avoid fatiguing children with repetitive tasks. 

Two online surveys were developed in Qualtrics \cite{qualtrics}. The first was completed by 6 participants, and the second by 4 new participants, all recruited through convenience sampling. Although descriptive, these surveys offered preliminary insights intended to guide stimulus selection.

For the first survey, ten game-phase sentences representative in length and structure were selected. Three versions of each sentence were created: unmodified, pitch increased by 20 Hz, and pitch decreased by 20 Hz. This manipulation is above the perceptual threshold for pitch differences \cite{jqzs17} but small enough to prevent artifacts \cite{app10}. The participants rated the calmness of all 30 items on a 10-point scale. Lower-pitched voices were indeed rated as more calming (M = 6.20, SD = 1.92) than unmodified (M = 5.66, SD = 1.66) and higher-pitched voices (M = 5.12, SD = 1.91), but the differences between conditions were not significant. 

Reexamination of the stimuli revealed an unintended intensity variation, possibly introduced by speech manipulations. A second survey using intensity-normalized stimuli (70dB, using \cite{w20}) was implemented to improve reliability. Additionally, to provide a comparison for pitch as a cue for stress, two conditions were added: a 10\% increase and decrease in speaking-rate (\cite{sjkkmd23}). The experimental procedure remained unchanged, but to reduce the length of the task, the stimuli were reduced to six sentences. Speaking rate manipulations affected perceived calmness to a similar extent as pitch manipulations as higher rates were judged to be less calming (M = 3.75, SD = 1.33), and the difference between the most and least calming speaking‑rate conditions was approximately one point on the scale. Notably, the unmodified voice was perceived as slightly more calming (M = 4.75, SD = 2.29) than the slowed version (M = 4.60, SD = 1.54), suggesting that bidirectional speaking‑rate adjustments may have inconsistent or non‑linear effects on perceived calmness (see \cite{sjkkmd23}). In contrast, pitch manipulations showed a clear and consistent pattern in the second survey as well: the lower‑pitched voice was rated as the most calming (M = 5.29, SD = 2.52), followed by the unmodified version (M = 4.75, SD = 2.29), and the higher‑pitched version (M = 4.42, SD = 2.04).

Both surveys provide evidence for a modest but linear effect of pitch on perceived calmness in synthetic speech, reinforcing the literature identifying pitch as a facilitator for stress regulation in natural speech. Based on these findings, a two-way manipulation of 35 Hz was used in the robot experiment to enhance perceptual salience and maximize potential impact while avoiding artifacts. 

\subsubsection{The Timed LEGO Building Task}
The game phase consisted of a timed LEGO-building task, based on prior work showing that time limits in such activities increase stress in child participants \cite{lslyhb25}, \cite{lklca16}, \cite{sqkd18}. LEGO set 40468 (Yellow Taxi) was used because it was age-appropriate for the 8-12-year-old participants, moderately challenging, and not marketed to a specific gender. 

\begin{figure}
    \centering
    \includegraphics[width=1\linewidth]{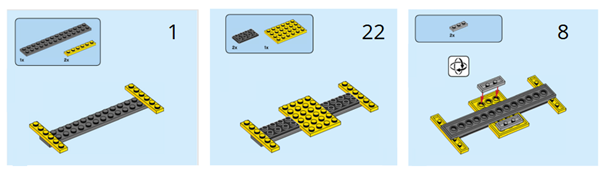}
    \caption{Example of renumbered LEGO instruction pages 1, 2, 3 from set 40468}
    \label{fig:LEGO}
\end{figure}

The set consisted of 124 bricks and 38 instruction steps. To create a guiding role for the robot, the original sequential step numbers were replaced with unique random numbers from 1 to 38, producing a non-sequential instruction order. As shown in Fig~\ref{fig:LEGO}, the original sequence (e.g., 1, 2, 3) was reordered into a randomized sequence (e.g., 1, 22, 8). This increased the task complexity and required participants to rely on the robot to identify the correct instruction page.

The renumbering was implemented by editing the digital instruction booklet in PowerPoint. Modified pages were then printed, cut, and laminated. 

Of the 38 steps, the first 20 were used in the experiment to create a task that was intentionally difficult to complete within the 10-minute limit. Participants were reminded of the time limit by a digital timer (from Basisonderwijs Online) displayed on a laptop. A bell sound signaled when the allotted time had expired.

\subsection{Procedure}
Participants were tested individually in quiet, familiar environments (care center or home). The experimenter explained the procedure and ensured the participant felt comfortable before starting and remained present on the other side of the room to monitor the session.

In the introduction phase, the robot introduced itself and engaged in casual conversation (e.g., asking about favorite games and LEGO). After the five-minute interaction, the robot entered a resting state and the experimenter administered Questionnaire 1.

After the child completed the questionnaire, the experimental conditions took effect in the game phase (see \ref{sec:expdes}). The task proceeded as follows: participants started with the first step of the LEGO set and started building once the timer was activated. After completion of each step, the participants consulted the robot for the next step number. Participants located the corresponding renumbered page and continued building. This continued until all steps were completed or the ten minute time limit had passed. At that point, Questionnaire 2 was administered. The robot then thanked the participant and ended the interaction.

\section{Results}
Statistical analyses were performed using JASP (version 0.19.3) \cite{jasp2024} and RStudio (version 2024.9.1.394) \cite{t25}. Additional packages were used: 'dplyr' package \cite{wfhmv24} for data clean-up, 'ggplot2' \cite{w16} for visualization, and ‘car’ \cite{fw19} and ‘lmtest’ \cite{hzfc99} for assumption checks. First, the validity of the measurement instrument and success of the game task in inducing stress were assessed. Then the primary analysis was conducted consisting of linear regression analyses examining the effect of the independent variable, robot voice (higher-pitched or lower-pitched), on the dependent variable, stress (CAM-S score).

\subsubsection{The Stress Measurement Questionnaires}
\begin{figure}
    \centering
    \includegraphics[width=1\linewidth]{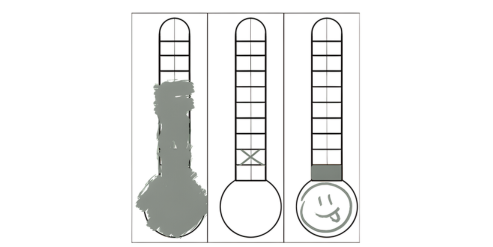}
    \caption{Examples of different coloring styles in the questionnaires}
    \label{fig:Termometers}
\end{figure}

The Participants completed a visual analog stress questionnaire before and after the game phase using a thermometer-style scale ranging from 0 to 10. Although response styles varied, from detailed coloring to simple marks and creative illustrations, the instrument yielded standardized numerical scores Fig~\ref{fig:Termometers}.

Both questionnaires consisted of 2 items (see section~\ref{sec:expdes}). The internal consistency between these items was strong, indicated by high Cronbach's alpha values for both the initial (α = .77) and second questionnaire (α = .97). Because of this, further analysis was conducted using only Item 1 (the original CAM-S measure \cite{ekmh13}).

\subsubsection{Stress Induction in the Game Phase}
The game phase was designed to induce stress by presenting a time-constrained task. Only eight of the 27 participants reached the final step. Participants on average reached step 16 (M = 16.3 steps, SD = 3.88), but there was a lot of variation with some advancing only to step 7. This could suggest that the task effectively introduced time pressure and performance uncertainty. Further analysis compared stress scores between the introduction and game phases. Participants reported higher stress during the timed game phase (M = 3.30, SD = 2.95) than during the introduction (M = 2.26, SD = 1.95), although stress scores remained low. A paired-samples t-test revealed a nonsignificant trend towards increased stress in the game phase (t(26) = 1.74, p = .09), suggesting that the task induced stress to some extent.\footnote{To further examine potential effects of the game session, a paired-samples t-test was also conducted on responses to Item 2 of the questionnaires. This revealed a significant effect (t (26) = 2.17, p = .04). Taken together with the trend observed in Item 1, this provides evidence that the game session was more stressful than the introduction.}

\subsubsection{The Effect of Robot Voice on Stress}
Stress scores were analyzed for two experimental conditions that were implemented in the game phase. For Condition 1, the robot used a higher pitch, in Condition 2 a lower pitch. Assumptions for linear regression were evaluated before analysis and indicate no evidence of nonlinearity, autocorrelation (D = 1.84, p = .79), heteroskedasticity (BP = 0.83, p = .36). These results confirmed that the regression analysis was appropriate.

\begin{figure}
    \centering
    \includegraphics[width=1\linewidth]{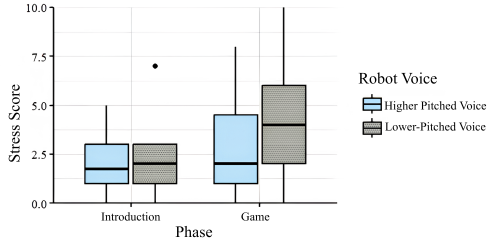}
    \caption{Mean stress scores during the introduction and game phases by condition}
    \label{fig:VoiceonStress}
\end{figure}

In the introduction phase, stress scores were slightly higher in Group 2 (M = 2.50, SD = 2.29) than in Group 1 (M = 2.04, SD = 1.39). A regression analysis revealed that this effect was insignificant (F(1, 25) = 0.41, p = 0.53), indicating that there were no meaningful differences between the groups prior to voice manipulation (see Fig~\ref{fig:VoiceonStress}). 

Contrary to expectations, Condition 2 elicited higher stress scores (M = 4.19, SD = 3.26) compared to Condition 1 (M = 2.71, SD = 2.62). Regardless, the effect was again not statistically significant, accounting for only 6.4\% of the variance in stress scores (F(1, 25) = 1.70, p = .20).\footnote{An exploratory analysis was conducted using the data from Item 2, this revealed no significant baseline effects (R² = .05, F(1, 25) = 1.31, p = .26) and no significant effects in the game session (R² = 0.7, F(1, 25) = 1.83, p = .19).}

\section{Discussion}
This study examined whether robot voice design, specifically pitch manipulations, can support stress regulation in children. Based on evidence from human communication showing that lower pitch is associated with reduced stress and greater emotional stability \cite{kl06}, \cite{xbviagn14}, \cite{btmrspf10}, as well as developmental findings demonstrating children's sensitivity to emotional prosody \cite{grossmann2010}, \cite{quam2011}, we hypothesized that using a lower-pitched robot voice in a stress inducing task would elicit lower self-reported stress. The results did not support this hypothesis, suggesting that the stress-modulating functions of pitch identified in human speech may not directly yield the same effects in synthetic robot voices in this context.

The sample size (N = 27) met the requirements of the power analysis, but limited the ability to detect subtle effects or explore contextual variables, such as differences between the home and care center testing environments. To minimize the effect of these variables, participants were tested in a familiar environment in a quiet room with the same experimental setup. Nevertheless, the testing environment may still have affected children’s comfort levels and stress responses. Future studies should consider recruiting larger samples or controlled environments to isolate such effects.

Stress was measured using an adapted two-item version of the CAM-S \cite{ekmh13}, with Item 1 assessing overall post-phase stress and Item 2 targeting phase-induced stress. Item 2 used less self-referential phrasing, attempting to mitigate social desirability biases (see \cite{se59}). High internal consistency was observed between the two items, indicating that they measured the same underlying construct and supporting the reliability of the adapted measure. However, the two items possibly diverged in their sensitivity to phase-related stress. For the comparison between stress scores reported in the introduction and game phase, the data from Item 1 showed a nonsignificant trend towards increased stress in the game phase, while Item 2 revealed a significant effect. It could also be that due to the task-focused (instead of self-focused) framing of Item 2 participants felt less hesitant to report negative emotions \cite{se59}. Moreover, participants were consistently presented first with Item 1. It could also be that this question was influenced by baseline stress or personality traits and served as a reference point for subsequent responses to Item 2.

Based on the observed increase in stress scores during the game phase, it can be concluded that the LEGO-building activity was moderately effective in inducing stress. On average, participants reported stress levels approximately one point higher in the game phase compared to the introduction. This pattern aligns with previous findings showing that introducing a time limit in LEGO tasks can be a stressor for children \cite{lslyhb25}, \cite{lklca16}, \cite{sqkd18}. Nevertheless, stress levels remained arguably low in both phases (2.2 out of 10 in the introduction; 3.3 in the game). This may reflect the low-stakes nature of the task, as the participants did not experience negative or positive consequences depending on the completion status. Previous studies (e.g., \cite{lslyhb25}) have shown that introducing reward or competition elements can heighten stress. Incorporating such elements into future experiment design can enhance the strength and ecological validity of stress induction. Moreover, the relatively low stress levels observed across conditions may be attributable to factors beyond the task design. One possible explanation could be a novelty effect \cite{flkalsntgmassdk22}, where the excitement of interacting with a robot may have overshadowed the stress-inducing aspects of the game. The current study attempted to mitigate this effect by showing the robot to the children prior to the experiment day and including a familiarization in the experiment but it was not examined if this was successful.

These limitations highlight the challenges of relying solely on self-report measures, particularly with children who may struggle to accurately assess or articulate their stress levels. Future research should consider the integration of physiological measures such as heart rate variability to provide objective data \cite{vlfsb24}, \cite{lslyhb25}. However, this approach presents practical and ethical challenges, especially with younger participants. Commercial devices may be ill-fitting or uncomfortable, and more accurate alternatives, such as chest bands, require assistance, potentially introducing discomfort or disrupting the experimental setting. Involving caregivers in the use of these devices \cite{lslyhb25} could be a viable solution, although it adds logistical complexity.

Furthermore, contrary to literature indicating that a lower vocal pitch can reduce stress \cite{btmrspf10}, \cite{kl06}, \cite{tlcl22}, the present study found no evidence that lowering a robot's pitch modulated children's stress levels. This suggests that the stress-reducing effects of pitch observed in human speech may not transfer directly to synthetic voices. 

Notably, the preliminary exploration suggested that lower-pitched synthetic voices may be perceived as less stressed, consistent with human voice cues. The absence of an effect in the main experiment could therefore suggest that this perceptual interpretation in synthetic voices may not translate into a direct effect on stress levels. In the main experiment, the embodiment of the robot also introduced additional social cues, such as gaze, posture and facial expressions. These cues may have overshadowed the prosodic manipulations or created a mismatch between the robot's appearance and its altered voice. For instance, the cute design of the robot may have made a lower-pitched voice seem incongruent and therefore less effective as a calming cue. This suggests a need to compare embodied and disembodied voice presentations in future research.

The lack of significant effects from pitch manipulation could also be explained by the fact that pitch modulation in natural interaction is a dynamic process. For example, while overall stress reduction can be facilitated by implementing lower voice pitch \cite{btmrspf10}, \cite{kl06}, \cite{tlcl22}, experienced therapists have been shown to strategically increase pitch in moments of high arousal to signal engagement \cite{galhjsh22}. This suggests that stable pitch manipulations, as implemented in the present study may not reflect how pitch modulation is naturally used throughout a session. Moreover, pitch rarely operates in isolation. Facilitating stress regulation through speech typically involves a combination of adjustments in, for example, speaking rate, intensity, and lexical content. Future studies should therefore explore multi-modal synthetic speech manipulations to better simulate emotionally supportive speech in real life. 

An alternative explanation relates to the differing magnitudes of pitch manipulation across the two studies. The validation study employed relatively subtle adjustments (±20Hz), whereas the main experiment used a broader manipulation (±35Hz). Although such increases in pitch range are not typically associated with adverse perceptual effects, unlike speaking‑rate manipulations, for example \cite{flkalsntgmassdk22} (see \ref{sec:robvoiceman}), the perceptual consequences of an expanded pitch range have not been systematically evaluated. It is therefore possible that this unexamined factor contributed to the null findings.

The age of the participants may also have played a role. The validation study involved adult proxies, whereas the main experiment was aimed at children. While similarities in emotional prosody recognition have been observed between adults and children (see \ref{sec:pitchstressac}), this ability continues to develop throughout \cite{flglcgg22}. Adults are already known to perceive emotional prosody in synthetic voices less effectively than in human speech \cite{crpgmz20}, suggesting that children may face even greater challenges in this domain. Further research is needed to explore how people of different ages perceive emotional cues in robot voices.

\section{Conclusion}
In conclusion, the present study did not find evidence that a lower pitch level in robot voices is beneficial for stress relief, as has been observed in human communication. Although a preliminary validation study showed that lower-pitched robot voices were perceived by adults as more calming, this effect did not translate to measurable stress reduction in the main experiment with child participants. Differences in the perception of emotional prosody between adults and children may explain this discrepancy.

The LEGO building task with Zenbo Junior II successfully induced mild stress, although overall stress levels remained low, possibly due to the novelty of the robot or the absence of real consequences. These findings suggest that pitch alone may be insufficient to influence stress in child-robot interactions. Future research should explore additional prosodic features and contextual factors, such as competition or prior robot exposure, to better understand and design effective robot-based stress relief interventions for children.

\section{Acknowledgments}This research was supported by a VICI grant from the Dutch Research Council (NWO) awarded to A. Chen (grant number VI.C.201.109).

\bibliographystyle{IEEEtran}
\bibliography{bibtex.bib}

\end{document}